\documentclass{ifacconf}

\usepackage[dvipsnames]{xcolor}

\usepackage{amsmath,amssymb,mathtools,bbm,dsfont,stmaryrd,aligned-overset,physics}
\allowdisplaybreaks

\newtheorem{theorem}{Theorem}[section]
\newtheorem{lemma}[theorem]{Lemma}
\newtheorem{corollary}[theorem]{Corollary}
\newtheorem{proposition}[theorem]{Proposition}
\theoremstyle{definition}

\newtheorem{assumption}{Assumption}[section]
\newtheorem{remark}{Remark}[section]
\theoremstyle{definition}

\usepackage{etoolbox}
\newbool{booltrue} 
\booltrue{booltrue}
\newbool{boolfalse} 
\boolfalse{boolfalse}

\usepackage[acronym]{glossaries}
\glsdisablehyper

\usepackage{enumerate,graphicx,epstopdf}
\usepackage{natbib}        
\usepackage{array}
\usepackage{algpseudocode,algorithm,algorithmicx}
\usepackage{booktabs}
\usepackage{lipsum}
\usepackage[deletedmarkup=sout,authormarkup=superscript]{changes}
\usepackage[normalem]{ulem}
\usepackage{comment}
\usepackage{tikz,pgfplots}
\usepackage{enumitem}
\usepgfplotslibrary{fillbetween,groupplots}
\pgfplotsset{compat=1.17}
\pgfplotsset{every axis/.append style={
                    label style={font=\normalsize},
                    tick label style={font=\small},
                    legend style={font=\small}
                    }}

\newcommand{\mc} {\mathcal}
\newcommand{\bb} {\mathbb}

\newcommand{\naturals}                       {\mathbb{N}}
\newcommand{\reals}                          {\mathbb{R}}

\DeclareMathOperator{\Id}                    {Id}

\DeclareMathOperator{\proj}                  {proj}

\newcommand{\onotation}[2][]                 {\ifstrequal{#1}{adapt}{o\left(#2\right)}{o(#2)}}

\DeclareMathOperator*{\argmin}               {arg\,min}

\newcommand{\innerProduct}[2]                {\left\langle #1,#2\right\rangle}

\newcommand{\transpose}[1]                   {{#1}^\top}

\let\var\relax
\DeclareMathOperator{\var}{Var}

\newcommand{\expectedValue}[2]               {\mathbb{E}^{#1}\left[#2\right]}
\newcommand{\variance}[2]                    {\var^{#1}\left[#2\right]}

\newcommand{\diracMeasure}[1]                {\delta_{#1}}

\newcommand{\support}[1]                     {\mathrm{supp}(#1)}

\newcommand{\Pp}[2]                          {\mathcal{P}_{#1}(#2)}

\newcommand{\wassersteinDistance}[3]         {W_{#1}(#2,#3)}

\newcommand{\setPlans}[2]                    {\Gamma(#1,#2)}
\newcommand{\setOptimalPlans}[2]             {\Gamma_o(#1,#2)}

\newcommand{\pushforward}[1]                 {#1_{\#}}

\renewcommand{\d}[0]                         {\mathrm{d}}

\renewcommand{\gradient}[1]                  {\nabla_{#1}}

\newcommand{\Lp}[2]                          {L^{#1}\ifstrempty{#2}{}{(#2)}}

\newcommand{\LpNorm}[3]                      {\norm{#1}_{L^{#2}(#3)}}

\newbool{finalsubmission} 
\boolfalse{finalsubmission} 

\definechangesauthor[name={Efe}, color=NavyBlue]{EB}
\definechangesauthor[name={Dominic}, color=RedViolet]{DLM}
\definechangesauthor[name={Nicolas}, color=OliveGreen]{NL}

\newacronym{acr:pdm}{PdM}{Predictive Maintenance}
\newacronym{acr:rls}{RLS}{Recursive Least Squares}

\begin{document}
\begin{frontmatter}


\title{Stochastic Wasserstein Gradient Flows using Streaming Data with an Application in Predictive Maintenance}


\thanks[footnoteinfo]{e-mails: \texttt{\{lnicolas,ebalta,dorfler\}@ethz.ch}, \\ \texttt{dliaomcp@mech.ubc.ca}. This research is supported by the Swiss National Science Foundation through NCCR Automation (Grant Number 180545).}

\author[ETH]{Nicolas Lanzetti} 
\author[ETH]{Efe C. Balta} 
\author[UBC]{Dominic Liao-McPherson} 
\author[ETH]{Florian Dörfler}

\address[ETH]{Automatic Control Laboratory, ETH Zürich, Physikstrasse 3, 8092 Zürich, Switzerland.}
\address[UBC]{University of British Columbia Department of Mechanical Engineering, 2054-6250 Applied Science Lane, \\Vancouver, BC V6T 1Z4, Canada.} 

\begin{abstract}                
We study estimation problems in safety-critical applications with streaming data. Since estimation problems can be posed as optimization problems in the probability space, we devise a stochastic projected Wasserstein gradient flow that keeps track of the belief of the estimated quantity and can consume samples from online data.
We show the convergence properties of our algorithm. Our analysis combines recent advances in the Wasserstein space and its differential structure with more classical stochastic gradient descent. 
We apply our methodology for predictive maintenance of safety-critical processes: Our approach is shown to lead to superior performance when compared to classical least squares, enabling, among others, improved robustness for decision-making. 
\end{abstract}


\end{frontmatter}

\section{Introduction}\label{sec:introduction}

Providing performance guarantees for parameter estimation algorithms operating with streaming data is a key challenge when developing methods for safety-critical applications across various domains of engineering and data science. Ideally, one should be able to (i) efficiently handle streaming data in real time, without resorting to computationally expensive one-shot numerical routines, and (ii) rigorously quantify the uncertainty related to the estimated quantity. In this paper, we focus on probabilistic approaches to uncertainty quantification rather than set-based ones~\citep{combettes1993foundations}.

A prominent approach for parameter estimation with streaming data is~\gls{acr:rls}. In~\gls{acr:rls}, the online solution is obtained by ``updating'' the previous solution with the latest measurement. \gls{acr:rls}  avoids the need to store and invert large data matrices and provides probabilistic guarantees on its estimate when the process is linear and all distributions are Gaussian.  \gls{acr:rls} is a special case of Bayes filter (see e.g., \cite{sarkka2013bayesian,sullivan2015introduction}), whose many variants (e.g., particle filters, extended Kalman filters, etc.) are the dominant approaches for inference using non-Gaussian distributions. Bayes filter is powerful but inflexible, it can be challenging to integrate side information and it can be difficult to implement due to the need to compute high-dimensional integrals.

In this work, we propose a different approach based on the theory of optimization in Wasserstein probability spaces~\citep{jordan1998variational,ambrosio2005gradient,lanzetti2022first}. We pose the parameter estimation problem as an optimization problem in the probability space and devise a stochastic projected gradient flow to iteratively compute its optimal solution using samples obtained from streaming data. Our approach maintains and iteratively improves an estimate of the probability measure of the estimated quantities and does not require a-priori assumptions on the probability measures (e.g., Gaussianity), but rather works in the space of all probability measures with finite second moment. Our proposed framework is more flexible than the Bayes filter in the sense that its intuitive to add side information e.g., constraints on the support of the final distribution or on the variance (indeed, it can be used to recover maximum likelihood estimator for stochastic least squares problems \citep{rigollet2018entropic}).

A motivating application of interest is \gls{acr:pdm}, where the goal is to efficiently maintain a safety-critical process (e.g., with minimal interruption) before an unsafe event occurs (see~\cite{pech2021predictive} for a recent survey of results). 
The \gls{acr:pdm} problem is challenging from an online algorithmic perspective since, in practice, there is often only historical data on the nominal operation and little or no data on the unsafe operation. 
Moreover, the problem calls for careful risk analysis: Too conservative decisions impact performance and efficiency, while unsafe events, if they occur, might lead to catastrophic failures.
The current state-of-the-art consists of rule-based methods and estimation strategies that rely on predetermined distribution models \citep{hu2020predictive,kanso2022remaining}.
With our work, we learn the model of the underlying process without a-priori assumptions on its probability measure to improve the overall performance by reducing conservativeness.

Our contributions are twofold. First, we propose a novel stochastic projected gradient flow for optimization in the probability space that operates on streaming data and study its convergence properties. Our analysis combines tools from optimal transport and differential calculus in the probability space with more classical projected stochastic gradient descent. We prove that similar to the Euclidean setting, our scheme yields convergence to a ball around the optimal solution.
Second, we apply our scheme to the predictive maintenance of the damping ratio of a second-order system and demonstrate improved performance relative to a classical least-squares approach.

\section{Background}\label{sec:background}

In this section, we briefly review our notation, basics of measure theory and optimal transport, geodesic convexity, and Wasserstein gradients. For more details, we refer the reader to~\cite{villani2009optimal,ambrosio2005gradient,santambrogio2015optimal,lanzetti2022first}.
\ifbool{finalsubmission}{All proofs of our theoretic results are relegated to the online extended version of this paper~\citep{lanzetti2023stochastic}.}{}

\subsubsection{Notation}
We consider the Euclidean space $\reals^d$, with the usual Euclidean norm $\norm{\cdot}$.
For a matrix $A\in\reals^{m\times n}$, we denote by $\sigma_\mathrm{min}(A)$ and by $\sigma_\mathrm{max}(A)$ its minimum and maximum singular value, respectively. If $m=n$, we use the notation $\lambda_\mathrm{min}(A)$ and by $\lambda_\mathrm{max}(A)$ for the minimum and maximum eigenvalue of $A$ and $\trace(A)$ for its trace.

\subsubsection{Basics in Measure Theory}
We denote by $\Pp{}{\reals^d}$ the space of (Borel) probability measures over $\reals^d$ and by $\Pp{2}{\reals^d}\coloneqq\{\mu\in\Pp{}{\reals^d}:\int_{\reals^d}\norm{x}^2\d\mu(x)<+\infty\}$ the space of probability measures with finite second moment.
We denote the Dirac measures at $x\in\reals^d$ by $\diracMeasure{x}$, defined by $\diracMeasure{x}(A)=1$ if and only if $x\in A$. 
We denote the support of a probability measure $\mu\in\Pp{2}{\reals^d}$ by $\support{\mu}\subset\reals^d$.
The pushforward of a measure $\mu\in\Pp{}{\reals^d}$ via a (Borel) map $T:\reals^d\to\reals^d$ is denoted by $\pushforward{T}\mu$ and defined by $(\pushforward{T}\mu)(B)=\mu(T^{-1}(B))$ for all $B\subset\reals^d$ Borel. For any $f:\reals^d\to\reals$, $\pushforward{T}\mu$-integrable it holds
\begin{equation*}
    \int_{\reals^d}f(x)\d(\pushforward{T}\mu)(x)
    =
    \int_{\reals^d}f(T(x))\d\mu(x).
\end{equation*}
A sequence of probability measures $(\mu_n)_{n\in\naturals}\subset\Pp{}{\reals^d}$ converges narrowly to $\mu\in\Pp{}{\reals^d}$ if $\int_{\reals^d}\phi(x)\d\mu(x)\to\int_{\reals^d}\phi(x)\d\mu(x)$ for all bounded continuous $\phi:\reals^d\to\reals$.

\subsubsection{Wasserstein distance}
The (type 2) Wasserstein distance between two probability measures $\mu, \nu\in\Pp{}{\reals^d}$ is 
\begin{equation*}
    \wassersteinDistance{2}{\mu}{\nu}
    \coloneqq\left(\min_{\gamma\in\setPlans{\mu}{\nu}}\int_{\reals^d\times\reals^d}\norm{x-y}^2\d\gamma(x,y)\right)^{\frac{1}{2}},
\end{equation*}
where $\setPlans{\mu}{\nu}$ is the set of transport plans, that is, of probability measures on $\reals^d\times\reals^d$ whose first marginal is $\mu$ and second marginal is $\nu$; i.e.,  $\setPlans{\mu}{\nu}=\{\gamma\in\Pp{}{\reals^d\times\reals^d}:\pushforward{(\proj_1)}\gamma=\mu,\pushforward{(\proj_2)}\gamma=\nu\}$ where $\proj_1$ and $\proj_2$ are projection operators (e.g., $\proj_1(x,y)=x$). We denote by $\setOptimalPlans{\mu}{\nu}$ the (non-empty) set of optimal couplings between $\mu$ and $\nu$. It is well-known that the Wasserstein distance is a distance on $\Pp{2}{\reals^d}$.

\subsubsection{Geodesic convexity}

Given $\mu_0\in\Pp{}{\reals^d}$ and $\mu_1\in\Pp{}{\reals^d}$, we define the geodesic between them by $\mu_t=\pushforward{((1-t)\proj_1+t\proj_2)}\gamma$, where $\gamma\in\setOptimalPlans{\mu_0}{\mu_1}$ is an optimal transport plan between $\mu_0$ and $\mu_1$. Since optimal transport plans are generally not unique, there might exist multiple geodesics between $\mu_0$ and $\mu_1$. 
Accordingly, a functional $J:\Pp{2}{\reals^d}\to\reals$ is $\alpha$-geodesically convex if for all $\mu_0,\mu_1\in\Pp{2}{\reals^d}$ there exists a geodesic $\mu_t$ so that $J(\mu_t)\leq (1-t)J(\mu_0)+tJ(\mu_1)-\frac{\alpha}{2}t(1-t)\wassersteinDistance{2}{\mu_0}{\mu_1}^2$. For instance, $\mu\mapsto \expectedValue{\mu}{V}$ is $\alpha$-geodesically convex if and only if $V:\reals^d\to\reals$ is $\alpha$-convex (i.e., convex with convexity parameter $\alpha$) and $\mu\mapsto\variance{\mu}{x_i}$ is geodesically convex (with $\alpha=0$), where $\variance{\mu}{x_i}$ denotes the variance of $x_i$.

\subsubsection{Wasserstein gradient}

A function $\gradient{\mu}J(\mu)\in\Lp{2}{\reals^d,\reals^d;\mu}$ is a Wasserstein gradient of a real-valued functional over the probability space $J:\Pp{2}{\reals^d}\to\reals$ if it approximates $J$ ``linearly''; i.e., for all $\gamma\in\setOptimalPlans{\mu}{\nu}$
\begin{equation*}
    J(\nu)-J(\mu)
    =
    \int_{\reals^d}\transpose{\gradient{\mu}J(\mu)(x)}(y-x)\d\gamma(x,y)+\onotation{\wassersteinDistance{2}{\mu}{\nu}},
\end{equation*}
where $\onotation{\wassersteinDistance{2}{\mu}{\nu}}$ denotes high-order term.
Wasserstein gradients are well-defined for many functionals of practical interest. In particular, we have $\gradient{\mu}\expectedValue{\mu}{V}=\gradient{}V$ for any smooth $V:\reals^d\to\reals$ with at most quadratic growth (i.e., the Wasserstein gradient of an expected value is simply the gradient of the function in the expected value) and $\gradient{}\variance{\mu}{x_i}=2(x_i-\expectedValue{\mu}{x_i})$. For the Wasserstein gradients of more functionals, we refer to~\cite{lanzetti2022first}. 

\section{Stochastic Projected Gradient Descent in Probability Spaces}\label{sec:gradient descent}
We construct our estimation method by encoding our objectives in an optimization problem and adapting a gradient descent algorithm to operate using samples from the system obtained with streaming data. Consider the optimization problem
\begin{equation}\label{eq:main optimization}
\begin{aligned}
    \inf_{\mu\in\Pp{2}{\reals^d}}
    &J(\mu)
    \\
    \text{s.t.}\quad &\support{\mu}\subset\Theta,
\end{aligned}
\end{equation}
we seek to minimize a real-valued lower semi-continuous\footnote{Here, lower semi-continuity is intended with respect to the convergence induced by the Wasserstein distance.} function $J:\Pp{2}{\reals^d}\to\reals$ over the probability space subject to a support constraint. The functional $J$ can encode standard expected values of real-valued quantities, but also other costs such as the variance, Wasserstein distance from a reference probability measure, or Kullback-Leibler divergence. 
We impose the following assumption on \eqref{eq:main optimization}:

\begin{assumption}\label{ass:theta}
The set $\Theta\subset\reals^d$ is closed and convex. 
\end{assumption}

Since we only have access to streaming data, we cannot evaluate $J$ and its Wasserstein gradient $\gradient{\mu}J$ exactly. Thus, we solve~\eqref{eq:main optimization} via a stochastic projected gradient descent, where at each iteration $k\in\naturals$ we have access to an unbiased noisy estimate of the Wasserstein gradient of $J$ and we leverage projections to enforce the support constraint. More specifically, our scheme reads
\begin{equation}\label{eq:stochastic wasserstein gradient flow}
\begin{aligned}
    \mu(k+1)
    &=
    \proj_{\text{supp}\subset\Theta}\left[
    \pushforward{(\Id-\tau\xi_k)}\mu(k)\right]
    \\
    \mu(0)&=\mu_0\in\Pp{2}{\reals^d},
\end{aligned}
\end{equation}
where $\Id$ is the identity map on $\reals^d$, $\xi_k$ is an unbiased estimate of the Wasserstein gradient, i.e., 
\begin{equation*}
\begin{aligned}
    \expectedValue{}{\xi_k} &=\gradient{\mu} J(\mu(k)),
\end{aligned}
\end{equation*}
$\tau\in\reals_{>0}$ is a step size, and $\proj_{\text{supp}\subset\Theta}\left[\cdot\right]$ denotes the projection (w.r.t. to the Wasserstein distance) onto the set of probability measures with support contained in $\Theta$. Later, we demonstrate how we construct our gradient estimate $\xi_k$ using streaming data.

We make the following assumption on our noisy gradients:
\begin{assumption}[Finite second moment]
\label{ass:finite variance}
The estimate of the gradient has bounded variance. In particular, there exists $\sigma>0$ and $C>0$ so that
\begin{equation*}
    \expectedValue{}{\norm{\xi}_{\Lp{2}{\reals^d,\reals^d; \mu}}^2} \leq \sigma^2 + C(J(\mu)-J(\mu^\ast)).
\end{equation*}
\end{assumption}
This assumption is mild: It stipulates that the second moment of the norm of the gradient at $\mu$ is controlled by the suboptimality of $\mu$. Whenever it is uniformly (in $\mu$) upper bounded, Assumption~\ref{ass:finite variance} holds trivially.

\subsection{Projections in the Wasserstein Space}
Our proposed algorithm includes a projection onto the set of probability measures with support in $\Theta$, denoted by $\proj_{\text{supp}\subset\Theta}\left[\cdot\right]$, which is defined by
\begin{equation*}
\begin{aligned}
    \proj_{\text{supp}\subset\Theta}[\mu]
    =
    \argmin\limits_{\bar\mu\in\Pp{2}{\reals}}~&\wassersteinDistance{2}{\mu}{\bar\mu}
    \\
    \text{s.t.}~& \support{\bar\mu}\subset\Theta.
\end{aligned}
\end{equation*}
Our next result states the projection of a probability measure onto the set of probability measures with support contained in $\Theta$ is (i) well-defined and (ii) results from pushforward of $\mu$ via the projection operator $\proj_\Theta:\reals^d\to\Theta$ on $\reals^d$. Intuitively, we can thus compute projections by ``projecting every (infinitesimal) particle of $\mu$ to $\Theta$'': 

\begin{proposition}[Projections]\label{prop:projection}
Let Assumption~\ref{ass:theta} hold.
Then, $\proj_{\Theta}\left[\cdot\right]$ is well-defined and for all $\mu\in\Pp{2}{\reals^d}$
\begin{equation}\label{eq:projection particles}
     \proj_{\text{supp}\subset\Theta}\left[
     \mu 
     \right]
     =
     \pushforward{\left(\proj_\Theta[\cdot]\right)}\mu. 
\end{equation}
\end{proposition}
Since every point $x\in\reals^d$ can be embedded to a probability measure $\diracMeasure{x}\in\Pp{2}{\reals^d}$, Assumption~\ref{ass:theta} is necessary for the existence of a unique projection. Indeed, if it fails to hold, then the projection operator is ill-defined even on $\reals^d$. In Proposition~\ref{prop:projection}, we show that it is also sufficient.

\subsection{Convergence Analysis}
We now study the convergence properties of the iteration~\eqref{eq:stochastic wasserstein gradient flow}. Similarly to Euclidean settings, the stochastic projected Wasserstein gradient descent~\eqref{eq:stochastic wasserstein gradient flow} converges to a (Wasserstein) ball centered at the optimal solution of~\eqref{eq:main optimization}: 

\begin{theorem}[Convergence]\label{thm:convergence}
Let $J:\Pp{2}{\reals^d}\to\reals$ be Wasserstein differentiable and $\alpha$-geodesically convex with convexity parameter $\alpha>0$, let Assumptions~\ref{ass:theta} and~\ref{ass:finite variance} hold, let $\mu^\ast\in\Pp{2}{\reals^d}$ be the optimal solution of~\eqref{eq:main optimization}, and let $\tau\in(0,\min\{1/\alpha,2/C\})$.
Then, for all $k\in\naturals$
\begin{equation}\label{eq:theorem convergence finite}
\begin{aligned}
    &\expectedValue{\{\xi_j\}_{j=0}^k}{\wassersteinDistance{2}{\mu_{k+1}}{\mu^\ast}^2}
    \\
    &\hspace{1cm}\leq 
    (1-\tau\alpha)^k\left(\wassersteinDistance{2}{\mu_0}{\mu^\ast}^2-\frac{\tau\sigma^2}{\alpha}\right)+\frac{\tau\sigma^2}{\alpha},
\end{aligned}
\end{equation}
In particular,
\begin{enumerate}[label=\arabic*)]
    \item
    \begin{equation}\label{eq:theorem convergence asymptotic}
        \limsup_{k\to\infty}\expectedValue{\{\xi_j\}_{j=1}^k}{\wassersteinDistance{2}{\mu_{k+1}}{\mu^\ast}}
    \leq 
    \sqrt{\frac{\tau\sigma^2}{\alpha}},
    \end{equation}
    
    \item\label{item thm:mean}  
    \begin{equation}\label{eq:theorem convergence mean}
    \begin{aligned}
        \hspace{-0.6cm}
        \limsup_{k\to\infty}\:
        &\expectedValue{}{
        \sqrt{\norm{m_{k+1}-m^\ast}^2+d(S_{k+1},S^\ast)^2}
        }
        \\
        &\leq 
        \sqrt{\frac{\tau\sigma^2}{\alpha}};
    \end{aligned}
    \end{equation}
    where the expectation is taken w.r.t. $\{\xi_j\}_{j=0}^k$, $m_k$ and $m^\ast$ are the mean of $\mu_k$ and $\mu^\ast$, $S_k$ and $S^\ast$ are their covariance matrices, and $d$ is the Bures distance between symmetric positive semidefinite matrices:
    \begin{equation}\label{eq:bures metric}
        d(S_{k+1},S^\ast)\coloneqq \sqrt{\trace(S_{k+1} + S^\ast - 2(S_{k+1}^{1/2} S^\ast S_{k+1}^{1/2})^{1/2})};
    \end{equation}
    
    \item\label{item thm:lipschitz} for any $L$-Lipschitz continuous function $\varphi:\reals^d\to\reals$,
    \begin{align}
        \limsup_{k\to\infty}\,
        &\expectedValue{\{\xi_j\}_{j=0}^k}{
        \left|\LpNorm{\varphi}{2}{\reals^d,\reals,\mu}\!-\!\LpNorm{\varphi}{2}{\reals^d,\reals,\mu^\ast}\right| 
        }
        \notag \\
        &\leq L\sqrt{\frac{\tau\sigma^2}{\alpha}}, \label{eq:theorem convergence lipschitz}
    \end{align}
    where the $L^2$ norm w.r.t. a probability measure $\nu$ is
    \begin{equation*}
        \LpNorm{\varphi}{2}{\reals^d,\reals,\nu}^2
        = \int_{\reals^d}\varphi(x)^2\d\nu(x).
    \end{equation*}
    
\end{enumerate}
\end{theorem}

We can specialize our results to the noise-free case ($\sigma=0$). This way, we recover the convergence properties of Wasserstein gradient flows (e.g., see~\cite{ambrosio2005gradient}): 

\begin{corollary}[Noise-free case]\label{cor:noise free}
Let $\sigma=0$. Then,
\begin{equation*}
    \lim_{k\to\infty}\wassersteinDistance{2}{\mu_k}{\mu^\ast}=0. 
\end{equation*}
\end{corollary}

Our results predicate convergence in expectation to a Wasserstein ball. This conclusion is in line with standard stochastic gradient descent; e.g., see~\cite{bottou2018optimization}.
Furthermore, the iterates not only converges to a Wasserstein ball but also provide practically relevant information if the generated solution $(\mu_k)_{k\in\naturals}$ is subsequently used for prediction or estimation purposes. In particular, one can deploy results in uncertainty propagation \citep{aolaritei2022uncertainty} to study the propagation of Wasserstein balls through prediction processes or leverage distributionally robust optimization to evaluate the \emph{worst-case} risk over Wasserstein balls~\citep{mohajerin2018data,blanchet2019quantifying,gao2022distributionally}.


\section{Estimation with Streaming Data} \label{sec:streaming-data}
We next specialize our scheme \eqref{eq:stochastic wasserstein gradient flow} to a meaningful special case and illustrate how it can be applied to problems with streaming data. 
We assume access to a stream of data $\{y_k\}$ generated by the process
\begin{equation}
    y_k = W\theta^* + w_k
\end{equation}
where $W\in\reals^{d\times d}$ is the known process matrix, $\theta^* \in \Theta$ is the parameter we would like to estimate, and $w_k$ is zero-mean uncorrelated noise with finite variance, probability measure $\nu$, and support $\mathcal{W}\subset\reals^d$. We pose the following parameter estimation problem 
\begin{align}
\inf_{\mu \in \Pp{2}{\Theta}}
 J(\mu)\coloneqq &\frac{1}{2}\int_{\mathcal W}\int_{\Theta}\norm{W\theta^\ast+w-W\theta}^2\d\mu(\theta)\d\nu(w)\notag \\* & + \frac{\rho}{2} \variance{\mu}{\theta_1+\ldots+\theta_d}\notag
\\*
\text{s.t.}\quad &\support{\mu}\subset\Theta, \label{eq:opt-problem}
\end{align}
where $\rho>0$. In words, $\mu$ is a probability measure over estimators $\theta$: We penalize the expected estimation error and a regularization term accounting for high variance, and we impose that the estimator lies in a set $\Theta$. If we could solve this problem (i.e., we had access to all data in a batch), then we would obtain a Dirac probability measure at the least squares estimator (provided that it lies in $\Theta$). Nonetheless, since we only have access to online streaming data, we need to compute the solution iteratively. 
We impose mild assumptions on the noise as well as some structure in the linear model $W$: 
\begin{assumption}[Noise]\label{assumption:example noise}
The noise $w$ is zero-mean and has finite variance $\sigma_w^2>0$.
\end{assumption}
\begin{assumption}[Invertible linear model]\label{assumption:example W}
The matrix $W$ is invertible. 
\end{assumption}
Intuitively, Assumption \ref{assumption:example noise} allows us to show that the second moment of the stochastic gradient is well-behaved (cf. Assumption~\ref{ass:finite variance}), which allows us to deploy Theorem~\ref{thm:convergence}. Assumption~\ref{assumption:example W}, instead, is required to ensure strong (geodesic) convexity of the objective function. 

Our estimation problem \eqref{eq:opt-problem} involves the unknown true parameters $\theta^*$ and cannot be solved directly. Instead, we derive a data-driven algorithm using $\{y_k\}$. To start, we show that the Wasserstein gradient of $J$ is well-defined, derive an expression for computing it, and show that Assumption~\ref{ass:finite variance}, required for Theorem~\ref{thm:convergence}, holds true: 
\begin{lemma}[Wasserstein gradients]\label{lemma:gradient}
Let Assumption~\ref{assumption:example noise} hold. 
The Wasserstein gradient of $J$ reads 
\begin{equation*}
\begin{aligned}
    \gradient{\mu}J(\mu)(\theta)
    &=
    W^\top W\left(\theta-\theta^\ast\right)
    +
    \rho\left(\theta-\expectedValue{\mu}{\theta}\right)
\end{aligned}
\end{equation*}
Moreover, 
\begin{equation}\label{eq:wasserstein gradient example}
    \xi(\theta,\hat y) = W^\top \left(W\theta -\hat y\right) +
    \rho\left(\theta-\expectedValue{\mu}{\theta}\right),
\end{equation} 
is an unbiased estimate of $\gradient{\mu}J(\mu)$ so that
\begin{equation}\label{eq:wasserstein gradient unbiased}
    \expectedValue{}{\xi}=\gradient{\mu} J(\mu) 
\end{equation}
and 
\begin{align}
    \expectedValue{}{\LpNorm{\xi}{2}{\reals^d,\reals^d;\mu}}
    \leq 
    &4\max\{\sigma_\mathrm{max}(W)^2,\rho\} \left(J(\mu)-J(\mu^\ast)\right)\notag  \\
    &+ 4\max\{\sigma_\mathrm{max}(W)^2,\rho\}\sigma^2. \label{eq:wasserstein gradient finite variance}
\end{align}
\end{lemma}

\begin{remark}\label{rem:noise gradient}
We can perturb $\xi$ via any function $f$ of $\theta$ and of a random parameter $\zeta$ satisfying $\expectedValue{}{f(\zeta,\theta)}=0$ for all $\theta\in\support{\mu}$, and still obtain an unbiased estimate of the Wasserstein gradient. This increases its second moment, which imposes a re-evaluation of the upper bound \eqref{eq:wasserstein gradient finite variance}.
\end{remark}

We solve \eqref{eq:opt-problem} using $\{y_k\}$ via the following stochastic gradient descent iteration:
\begin{equation}\label{eq:example gradient iterate}
\begin{aligned}
    \mu(k+1)
    &=
    \proj_{\text{supp}\subset\Theta}\left[
    \pushforward{(\Id-\tau\xi_k)}\mu(k)\right]
    \\
    &=
    \pushforward{\left(\proj_{\Theta}[
    \Id+\tau\xi_k 
    ]
    \right)}\mu(k),
\end{aligned}
\end{equation} 
where $\xi_k = \xi(\theta,y_k)$ is our streaming data based estimate of $\nabla_\mu J$.

Convergence of~\eqref{eq:example gradient iterate} follows directly from Theorem~\ref{thm:convergence}.

\begin{corollary}[Convergence]\label{cor:convergence example}
Let Assumptions~\ref{ass:theta}, \ref{assumption:example noise}, and \ref{assumption:example W} hold. 
Let $\tau\in (0,1/(2\max\{\sigma_\mathrm{max}(W)^2,\rho\}))$, and let $(\mu_k)_{k\in\naturals}\subset\Pp{2}{\reals^d}$ be the sequence generated by~\eqref{eq:example gradient iterate} and $\mu^\ast$ be the optimal solution. Then,
\begin{equation}
\begin{aligned}
    &\expectedValue{\{\xi_j\}_{j=0}^k}{\wassersteinDistance{2}{\mu_{k+1}}{\mu^\ast}^2}
    \\
    &\hspace{0.25cm}\leq 
    \begin{aligned}[t]
    &(1\!-\!\sigma_\mathrm{min}(W)^2\tau)^k\left(\wassersteinDistance{2}{\mu_0}{\mu^\ast}^2\!-\!\tau\eta\sigma_w^2\right)\!+\!\tau\eta\sigma_w^2,
    \end{aligned}
\end{aligned}
\end{equation}
where $\eta\coloneqq 4\max\{\sigma_\mathrm{max}(W)^2,\rho\}/\sigma_\mathrm{min}(W)^2$.
In particular:
\begin{enumerate}[label=\arabic*)]
    \item $\limsup_{k\to\infty}\expectedValue{\{\xi_j\}_{j=0}^k}{\wassersteinDistance{2}{\mu_{k+1}}{\mu^\ast}}
    \leq 
    \sigma_w\sqrt{\eta\tau};$
    
    \item with $m_k$ and $m^\ast$ being the mean of $\mu_k$ and $\mu^\ast$, respectively, and $S_k$ and $S^\ast$ being their covariance matrices, 
    \begin{equation*}
    \begin{aligned}
        \hspace{-0.6cm}
        \limsup_{k\to\infty}\:
        &\expectedValue{\{\xi_j\}_{j=0}^k}{
        \sqrt{\norm{m_{k+1}-m^\ast}^2+d(S_{k+1},S^\ast)^2}
        }
        \\
        &\leq 
        \sigma_w\sqrt{\eta\tau},
    \end{aligned}
    \end{equation*}
    where $d$ is the Bures distance defined in~\eqref{eq:bures metric};
    
    \item for any $L$-Lipschitz continuous function $\varphi:\reals^d\to\reals$,
    \begin{equation*}
    \begin{aligned}
        \limsup_{k\to\infty}\,
        &\expectedValue{\{\xi_j\}_{j=0}^k}{
        \left|\LpNorm{\varphi}{2}{\reals^d,\reals,\mu}-\LpNorm{\varphi}{2}{\reals^d,\reals,\mu^\ast}\right| 
        }
        \\
        &\leq L \sigma_w\sqrt{\eta\tau}.
    \end{aligned}
    \end{equation*}
\end{enumerate}
\end{corollary}

\section{Predictive Maintenance of the Damping Ratio}\label{sec:example}

Consider the second-order system
\begin{equation} \label{eq:second_order}
   \ddot{z} + a\dot z + b (z - r + \varepsilon) = 0,
\end{equation}
where $a,b\in \reals$ are parameters, $\varepsilon$ is measurement noise with bounded variance and $r\in \reals$ is a reference signal. 
Our goal is to monitor the damping ratio
\begin{equation}
  \zeta\coloneqq\frac{a}{2\sqrt{b}} 
\end{equation}
and ensure that it does not violate the safe lower bound $\zeta_{\min} \in \reals$. This leads to the following safe set for $(a,b)$:
\begin{equation*}
   Q_\mathrm{safe} =\left\{(a,b)\in\reals^2: \frac{a}{2\sqrt{b}} \geq \zeta_{\min} \right\}\subset\reals^2.
\end{equation*}

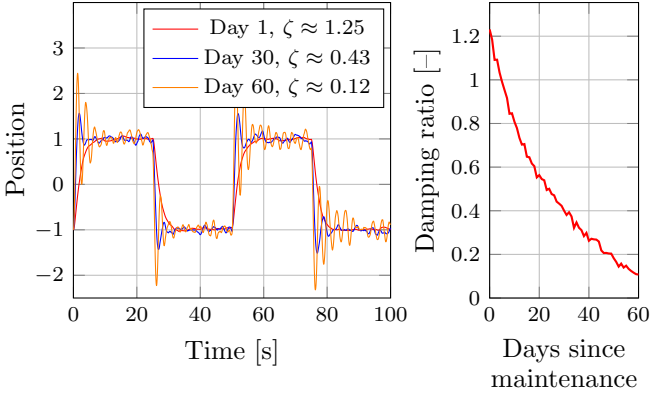
\begin{figure}[htbp]
   \centering
   \begin{tikzpicture}
   \begin{groupplot}[group style={group size=2 by 1,horizontal sep=1.3cm},
                     height=5.5cm]
   \nextgroupplot[width=0.65\columnwidth,xmin=0,xmax=100,
                  ymin=-2.5,ymax=4,
                  grid=both,
                  ytick={-3,-2,...,3},
                  ylabel={Position},xlabel={Time [s]},
                  ylabel style={yshift=-0.1cm}]
    \addplot[-,draw=red] table[x index=0,y index=1,col sep=comma] {figures/System/Responses.csv};
    \addlegendentry{Day 1, $\zeta\approx 1.25$}
    \addplot[-,draw=blue] table[x index=0,y index=2,col sep=comma] {figures/System/Responses.csv};
    \addlegendentry{Day 30, $\zeta\approx 0.43$}
    \addplot[-,draw=orange] table[x index=0,y index=3,col sep=comma] {figures/System/Responses.csv};
    \addlegendentry{Day 60, $\zeta\approx 0.12$}
    
   
   \nextgroupplot[width=0.4\columnwidth,xmin=0,xmax=60,ymin=0,
                  ylabel={Damping ratio [--]},
                  xlabel={Days since \\ maintenance},
                  ytick={0,0.2,...,1.4},
                  grid=both,
                  ylabel style={yshift=-0.1cm},
                  xlabel style={align=center}]
   \addplot[-,thick,draw=red] table[x index=0,y index=1,col sep=comma] {figures/System/Ratio.csv};
   \end{groupplot}
   \end{tikzpicture}
   \caption{As the damping ratio decays over time (right), the response of the system becomes increasingly oscillatory (left).}
   \label{fig:plant}
\end{figure}

The parameters $a$ and $b$ vary slowly with time according to the equation
\begin{equation} \label{eq:second_order_decay}
   y(t)
  =
  \begin{bmatrix}
    a(t) \\ b(t)
  \end{bmatrix}
  = \begin{bmatrix}
      a_0 - \lambda_1 t\\
      b_0 + \lambda_2 t
   \end{bmatrix},
\end{equation}
where $t\geq 0$ is the amount of time that has passed since the system was last maintained, $\theta = (\lambda_1,\lambda_2)\in\reals^2$ are unknown decay parameters, and  $a_0,b_0\in\reals$ are known constants. At time $t=0$, i.e., immediately after maintenance, the parameters $a$ and $b$ belong to the safe set; i.e., $(a_0,b_0)\in Q_\mathrm{safe}$. The coefficients $\lambda_1$ and $\lambda_2$ are positive (i.e., $\Theta=\reals_{\geq 0}^2$) so that $\zeta(t)$ decreases with time and the system will eventually exit the safe set. The behavior of the system as it decays is illustrated in Figure~\ref{fig:plant}.

We are interested in deciding when to perform maintenance on the system \eqref{eq:second_order}, which resets $t = 0$ and the parameters $a$ and $b$. Performing maintenance is expensive and it is desirable to do it as infrequently as possible while still ensuring that $a(t)$ and $b(t)$ remain in $Q_\mathrm{safe}$. Ideally we would always maintain the system at time
\begin{equation}\label{eq:tstar}
   t^\star = \sup\{t\geq 0: y(t) \in Q_\mathrm{safe}\}.
\end{equation}
However, in practice $\theta$ is unknown and we cannot directly measure $y$, hence we must infer $\theta$ from noisy data, use it to estimate $t^\star$, and account for the uncertainty in our estimation in our decision-making process.

\subsection{Estimation}\label{subsec:estimation}
The parameters $y$ cannot be directly measured but must be estimated based on trajectories of the system \eqref{eq:second_order}. For a fixed value of $t$ (remember that $a(t)$ and $b(t)$ vary slowly relative to the dynamics of \eqref{eq:second_order}) applying Euler discretization with a sampling period $\Delta t > 0$ to \eqref{eq:second_order} yields the discrete-time system
\begin{equation} \label{eq:discretized}
   x_{k+1} = \begin{bmatrix}
      1 & \Delta t\\
      -\Delta t b & 1 - \Delta t a
   \end{bmatrix} x_k + \begin{bmatrix}
      0 \\ \Delta t b
   \end{bmatrix} (r + \varepsilon_k),
\end{equation}
where the state is $x = (z,\dot{z})$, which we can rewrite as
\begin{equation*}
   x_{k+1} = A(y) x_k + B(y) (r_k + \varepsilon_k).
\end{equation*}
For sufficiently small $\Delta t$ and $a,b > 0$, \eqref{eq:second_order} is robustly stable about $z = r$. We then measure trajectories $\{\hat{x}_k,r_k\}_{k = 0}^N$ of \eqref{eq:discretized} and estimate $y(t)$ using the least-squares estimator
\begin{equation} \label{eq:estimator}
   \hat y(t) = \argmin_{y}\sum_{k=0}^{N-1}\Vert \hat{x}_{k+1} - A(y)\hat{x}_k - B(y)r_k\Vert ^2.
\end{equation}
The noise term $\varepsilon$ in \eqref{eq:discretized} introduces noise in the estimator \eqref{eq:estimator}. Thus, in practice, we obtain noisy measurements
\begin{equation*}
   \hat{y}(t) = \begin{bmatrix}
    a(t) \\ b(t)
  \end{bmatrix}
  = \begin{bmatrix}
      a_0 - \lambda_1 t\\
      b_0 + \lambda_2 t
   \end{bmatrix} + w(t)
\end{equation*}
where the noise term $w(t)\in \reals^2$ is uncorrelated in time and is assumed to have bounded variance. In this case, the noise $w$ results from the propagation of $\varepsilon$ through the argmin in~\eqref{eq:estimator} and so might not be not zero-mean.
In our case study, we generate trajectories of 100s with sampling time $\Delta t=0.001$s and suppose $\varepsilon_k$ is uniform on $[-3,+3]$.

\subsection{Probabilistic Predictive Maintenance}
To ensure robustness and careful decision-making, we adopt a probabilistic approach and encapsulate our belief about $\theta = (\lambda_1,\lambda_2)$ in a probability distribution $\mu \in \mc{P}_2(\reals^2)$ that will enable us to quantify our uncertainty about $\theta$. This opens the floor to stochastic and (distributionally) robust decision-making; e.g., with $\nu_t$ denoting the probability distribution of $y(t)$, we can use a chance constraint
\begin{equation*}
   t^\star = \sup\left\{t\geq 0: \bb{P}^{\nu_t}[Q_\mathrm{safe}]=\bb{P}^{x\sim\nu_t}[x\in Q_\mathrm{safe}] \geq 1-\alpha\right\}
\end{equation*}
for some confidence level $\alpha \in (0,1)$ or the mean prediction
\begin{equation*}
   t^\star = \sup\left\{t\geq 0: \expectedValue{\nu_t}{x}=\expectedValue{x\sim\nu_t}{x}\in Q_\mathrm{safe}\right\}.
\end{equation*}
Each day $k$, we obtain degradation data $\{\hat y_k,t_k\}$ from the system. To put our PDM problem \eqref{eq:second_order_decay} in the form of \eqref{eq:opt-problem} we consider the difference between two consecutive measurements, happening every $t_{k+1} - t_k = T>0$ time units, and obtain a new measurement function
\begin{equation*}
    \tilde y = y(t+T)-y(t) = 
    \begin{bmatrix}
    -\lambda_1 T \\ \lambda_2 T
    \end{bmatrix}
    +
    \begin{bmatrix}
    \tilde w_1 \\  \tilde w_2
    \end{bmatrix},
\end{equation*}
for which we have data $\{\hat y_{k+1}-\hat y_k\}$. Thus we have $W=\text{diag}(-T,T)$. The noise $\tilde w$ is zero-mean, since $\expectedValue{}{\tilde w_i}=\expectedValue{}{w_i(t+T)-w_{i}(t)}=0$, and has variance
\begin{equation*}
    \variance{}{\tilde w_i}=\variance{}{w_{i}(t+T)}+\variance{}{w_{i}(t)}\eqqcolon \sigma_w^2,
\end{equation*}
and the parameters are known to lie in the set $\Theta = \reals_{\geq0}^2$ which defines the support or $\mu$.

To obtain a practical implementation, we implement \eqref{eq:example gradient iterate} using particles, i.e., in the setting of probability measures that have finitely many samples of form $\mu(k)=\frac{1}{N}\sum_{i=1}^N\diracMeasure{x_i}$ with $\{x_i\}_{i=1}^N\subset\reals^d$. In this work, $N=1000$. In this case, the update equation for probability measures~\eqref{eq:example gradient iterate} simplifies to
\begin{equation}
    \mu(k+1)
    =
    \frac{1}{N}\sum_{i=1}^N\diracMeasure{\proj_{\Theta}[x_i-\tau\xi_k(x_i, y_k)]}.
\end{equation}
where $\tau>0$ is a step size and $\xi$ is the unbiased estimator of the Wasserstein gradient from Lemma~\ref{lemma:gradient}. That is, one can simply pushforward all ``particles'' of $\mu(k)$ and then project them individually to $\Theta$. It suffices therefore to keep track of the location $x_i\in\reals^d$ of every particle $i\in\{1,\ldots,N\}$. In particular, the update rule is then $x_i\mapsto\proj_{\Theta}[x_i-\tau\xi_k(x_i,y_k)]$, which can be evaluated in parallel for each particle $i$.
Accordingly, the computational complexity is the one of running parallel projected gradient descent iterations. As such, we are also subject to the usual trade-offs of gradient descent (e.g., optimality vs. number of iterations). In our case study, each gradient step takes on average 0.0025s (with standard deviation 0.0033s, Matlab implementation on a MacBook Pro with 2.3 GHz Quad-Core Intel Core i5).

\subsection{Numerical results}
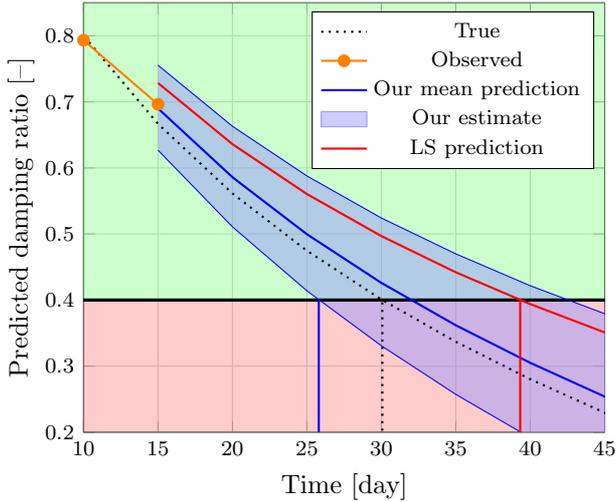
\begin{figure}[ht]
    \centering 
    \begin{tikzpicture}
    \begin{axis}[xmin=10,xmax=45,ymin=0.2,ymax=0.85,grid=both,
                 ytick={0.2,0.3,...,0.8},xtick={5,10,...,45},
                 xlabel={Time [day]},ylabel={Predicted damping ratio [--]},
                 ylabel style={yshift=-0.1cm},]
    \addplot[-,name path=safeOne,draw=black,very thick,forget plot] table[x index=0,y index=1,col sep=comma]  {figures/Confidence/thresholdTStarTrueOursLS.csv};
    \addplot[,name path=safeTwo,forget plot] coordinates {(10,2)(50,2)};
    \addplot[,name path=safeThree,forget plot] coordinates {(10,0)(50,0)};
    \addplot[green!40!white,fill opacity=0.5,forget plot] fill between[of=safeOne and safeTwo];
    \addplot[red!40!white,fill opacity=0.5,forget plot] fill between[of=safeOne and safeThree];
    
    \addplot[-,dotted,thick] table[col sep=comma]  {figures/Confidence/RatioTrue.csv};
    \addplot[-,draw=orange,thick,mark=*,mark options={fill=orange}] table[col sep=comma]  {figures/Confidence/RatioMeasured.csv};
    
    \addplot[-,draw=blue,thick] table[x index=0,y index=3,col sep=comma]  {figures/Confidence/RatioOurs.csv};
    \addplot[-,name path=A,draw=blue,forget plot] table[x index=0,y index=1,col sep=comma]  {figures/Confidence/RatioOurs.csv};
    \addplot[-,name path=B,draw=blue,forget plot] table[x index=0,y index=2,col sep=comma]  {figures/Confidence/RatioOurs.csv};
    \addplot[blue!40!white,fill opacity=0.5] fill between[of=A and B];
    
    \addplot[-,draw=red,thick] table[col sep=comma]  {figures/Confidence/RatioLS.csv};
    
    \addplot[-,dotted,draw=black,thick] table[x index=2,y index=5,col sep=comma]  {figures/Confidence/thresholdTStarTrueOursLS.csv};
    \addplot[-,draw=blue,thick] table[x index=3,y index=5,col sep=comma]  {figures/Confidence/thresholdTStarTrueOursLS.csv};
    \addplot[-,draw=red,thick] table[x index=4,y index=5,col sep=comma]  {figures/Confidence/thresholdTStarTrueOursLS.csv};

    \addlegendentry{True}
    \addlegendentry{Observed}
    \addlegendentry{Our mean prediction}
    \addlegendentry{Our estimate}
    \addlegendentry{LS prediction}
    
    \end{axis}
    \end{tikzpicture}

    \caption{ Prediction of the evolution of the damping ratio constructed at day 15, alongside with its standard least squares prediction and the true evolution of the damping ratio. For our estimate, we construct the confidence interval via the 10\% percentile and the 90\% percentile. The vertical lines highlight the intersection of the curves with the threshold $\zeta_\mathrm{min}=0.4$. The green area denotes safe operation ($\zeta\geq\zeta_\mathrm{min}$) and the red area denotes unsafe operation ($\zeta<\zeta_\mathrm{min}$).}
    \label{fig:prediction}
\end{figure}

\begin{figure}[ht]
    \centering
    \begin{tikzpicture}
    \begin{axis}[axis equal image,xmin=10,xmax=45,ymin=15,ymax=40,grid=both,
                 ytick={0,5,...,50},xtick={0,5,...,50},
                 xlabel={Time [day]},ylabel={Suggested maintenance time [day]},
                 ylabel style={yshift=-0.1cm},
                 legend pos=south east]
    \addplot[-,name path=safeOne,draw=black,very thick,forget plot] table[x index=1,y index=0,col sep=comma]  {figures/tStar/thresholdTStarTrueOursLS.csv};
    \addplot[,name path=safeTwo,forget plot] coordinates {(0,0)(0,50)};
    \addplot[,name path=safeThree,forget plot] coordinates {(50,0)(50,50)};
    \addplot[green!40!white,fill opacity=0.5,forget plot] fill between[of=safeOne and safeTwo];
    \addplot[red!40!white,fill opacity=0.5,forget plot] fill between[of=safeOne and safeThree];
    
    \addplot[-,very thick] table[col sep=comma]  {figures/tStar/tStarTrue.csv};
    \addplot[-,draw=blue,thick,mark=*,mark options={fill=blue}] table[col sep=comma]  {figures/tStar/tStarOurs.csv};
    \addplot[-,draw=red,thick,mark=*,mark options={fill=red}] table[col sep=comma]  {figures/tStar/tStarLS.csv};
    
    \addplot[-,dashed] coordinates {(0,0)(50,50)};

    \addlegendentry{True}
    \addlegendentry{Our estimate}
    \addlegendentry{LS estimate}
    \end{axis}
    \end{tikzpicture}
    
    \caption{Predicted maintenance time at each time step. Here, classical least squares fails to predict the maintenance time. Our algorithm, instead, robustly suggests to schedule maintenance a few days in advance. The green area denotes safe operation ($\zeta\geq\zeta_\mathrm{min}$) and the red area denotes unsafe operation ($\zeta<\zeta_\mathrm{min}$).}
    \label{fig:tstar}
\end{figure}
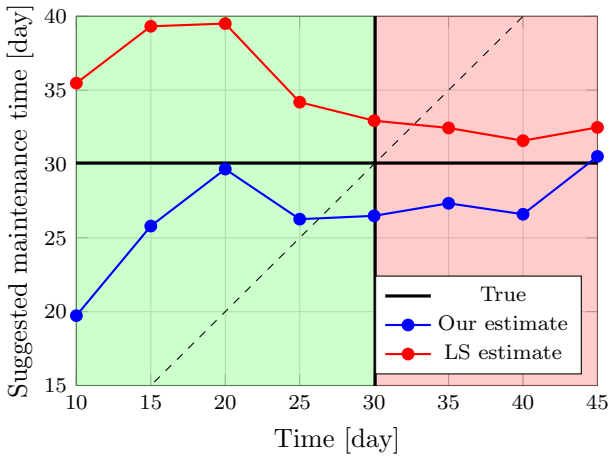

We use the true values $a_0=2.5, b_0=1$ (known) and $\lambda_1=2/60, \lambda_2=5/60$ (unknown). 
We initialize the particles' position via uniform sampling in $[0,8/60]^2\subset\reals^2$.
We weigh the variance with $\rho=0.1$ and consider $T=5$ days.
At each time step of the algorithm, (i) we collect an estimate of $y$ as described in Section~\ref{subsec:estimation}, (ii) we run one iterate of our gradient descent scheme, with $\tau$ sufficiently small and an additional zero-mean noise term (Gaussian with standard deviation 0.02) in the Wasserstein gradient (cf. Remark~\ref{rem:noise gradient}). 
Thereafter, we use the probability measure to construct a confidence interval for the damping, which can be used to schedule maintenance. For instance, Figure~\ref{fig:prediction} shows the confidence interval constructed at day 15, compared against a classical static least-squares estimate (for which there are no hyperparameters). This way, we can predict maintenance. 
We collect the predictive maintenance time at each iteration in Figure~\ref{fig:tstar}. 
As can be seen, our approach has superior performance than the classic least squares, as it readily enables robust decision-making, which consistently leads to safe estimates of the maintenance time. 

\section{Conclusions}\label{sec:conclusions}

In this work, we present a novel stochastic Wasserstein gradient flow method to efficiently perform estimation in probability spaces with streaming data.
Our formal results provide a convergence analysis of our online stochastic optimization method, which provides convergence to a ball around the optimal solution, similar to the standard Euclidean setting.
We illustrate the utility of the proposed method in an application of predictive maintenance to show the benefit over classical approaches such as simple least-squares with Gaussianity assumptions.
Overall, our method provides a flexible online estimation tool to estimate a rich set of processes without any assumptions on the model of the underlying distribution.
Future work will consider providing further results under relaxed settings such as non-strong convexity and applications for predictive maintenance of real-world physical examples.


\bibliography{ifacconf}          

\begin{thebibliography}{18}
\providecommand{\natexlab}[1]{#1}
\providecommand{\url}[1]{\texttt{#1}}
\providecommand{\urlprefix}{URL }
\expandafter\ifx\csname urlstyle\endcsname\relax
  \providecommand{\doi}[1]{doi:\discretionary{}{}{}#1}\else
  \providecommand{\doi}{doi:\discretionary{}{}{}\begingroup
  \urlstyle{rm}\Url}\fi

\bibitem[{Ambrosio et~al.(2005)Ambrosio, Gigli, and
  Savar{\'e}}]{ambrosio2005gradient}
Ambrosio, L., Gigli, N., and Savar{\'e}, G. (2005).
\newblock \emph{Gradient flows: in metric spaces and in the space of
  probability measures}.
\newblock Springer Science \& Business Media.

\bibitem[{Aolaritei et~al.(2022)Aolaritei, Lanzetti, Chen, and
  D{\"o}rfler}]{aolaritei2022uncertainty}
Aolaritei, L., Lanzetti, N., Chen, H., and D{\"o}rfler, F. (2022).
\newblock Uncertainty propagation via optimal transport ambiguity sets.
\newblock \emph{arXiv preprint arXiv:2205.00343}.

\bibitem[{Blanchet and Murthy(2019)}]{blanchet2019quantifying}
Blanchet, J. and Murthy, K. (2019).
\newblock Quantifying distributional model risk via optimal transport.
\newblock \emph{Mathematics of Operations Research}, 44(2), 565--600.

\bibitem[{Bottou et~al.(2018)Bottou, Curtis, and
  Nocedal}]{bottou2018optimization}
Bottou, L., Curtis, F.E., and Nocedal, J. (2018).
\newblock Optimization methods for large-scale machine learning.
\newblock \emph{Siam Review}, 60(2), 223--311.

\bibitem[{Combettes(1993)}]{combettes1993foundations}
Combettes, P.L. (1993).
\newblock The foundations of set theoretic estimation.
\newblock \emph{Proceedings of the IEEE}, 81(2), 182--208.

\bibitem[{Gao and Kleywegt(2022)}]{gao2022distributionally}
Gao, R. and Kleywegt, A. (2022).
\newblock Distributionally robust stochastic optimization with {W}asserstein
  distance.
\newblock \emph{Mathematics of Operations Research}.

\bibitem[{Gelbrich(1990)}]{gelbrich1990formula}
Gelbrich, M. (1990).
\newblock On a formula for the {L2} {W}asserstein metric between measures on
  {E}uclidean and {H}ilbert spaces.
\newblock \emph{Mathematische Nachrichten}, 147(1), 185--203.

\bibitem[{Hu and Chen(2020)}]{hu2020predictive}
Hu, J. and Chen, P. (2020).
\newblock Predictive maintenance of systems subject to hard failure based on
  proportional hazards model.
\newblock \emph{Reliability Engineering \& System Safety}, 196, 106707.

\bibitem[{Jordan et~al.(1998)Jordan, Kinderlehrer, and
  Otto}]{jordan1998variational}
Jordan, R., Kinderlehrer, D., and Otto, F. (1998).
\newblock The variational formulation of the {F}okker--{P}lanck equation.
\newblock \emph{SIAM journal on mathematical analysis}, 29(1), 1--17.

\bibitem[{Kanso et~al.(2022)Kanso, Jha, Galeotta, and
  Theilliol}]{kanso2022remaining}
Kanso, S., Jha, M.S., Galeotta, M., and Theilliol, D. (2022).
\newblock Remaining useful life prediction with uncertainty quantification of
  liquid propulsion rocket engine combustion chamber.
\newblock \emph{IFAC-PapersOnLine}, 55(6), 96--101.

\bibitem[{Lanzetti et~al.(2022)Lanzetti, Bolognani, and
  D{\"o}rfler}]{lanzetti2022first}
Lanzetti, N., Bolognani, S., and D{\"o}rfler, F. (2022).
\newblock First-order conditions for optimization in the {W}asserstein space.
\newblock \emph{arXiv preprint arXiv:2209.12197}.

\bibitem[{Mohajerin~Esfahani and Kuhn(2018)}]{mohajerin2018data}
Mohajerin~Esfahani, P. and Kuhn, D. (2018).
\newblock Data-driven distributionally robust optimization using the
  {W}asserstein metric: Performance guarantees and tractable reformulations.
\newblock \emph{Mathematical Programming}, 171(1), 115--166.

\bibitem[{Pech et~al.(2021)Pech, Vrchota, and
  Bedn{\'a}{\v{r}}}]{pech2021predictive}
Pech, M., Vrchota, J., and Bedn{\'a}{\v{r}}, J. (2021).
\newblock Predictive maintenance and intelligent sensors in smart factory.
\newblock \emph{Sensors}, 21(4), 1470.

\bibitem[{Rigollet and Weed(2018)}]{rigollet2018entropic}
Rigollet, P. and Weed, J. (2018).
\newblock Entropic optimal transport is maximum-likelihood deconvolution.
\newblock \emph{Comptes Rendus Mathematique}, 356(11-12), 1228--1235.

\bibitem[{Santambrogio(2015)}]{santambrogio2015optimal}
Santambrogio, F. (2015).
\newblock Optimal transport for applied mathematicians.
\newblock \emph{Birk{\"a}user, NY}, 55(58-63), 94.

\bibitem[{S{\"a}rkk{\"a}(2013)}]{sarkka2013bayesian}
S{\"a}rkk{\"a}, S. (2013).
\newblock \emph{Bayesian filtering and smoothing}.
\newblock 3. Cambridge university press.

\bibitem[{Sullivan(2015)}]{sullivan2015introduction}
Sullivan, T.J. (2015).
\newblock \emph{Introduction to uncertainty quantification}, volume~63.
\newblock Springer.

\bibitem[{Villani(2009)}]{villani2009optimal}
Villani, C. (2009).
\newblock \emph{Optimal transport: old and new}, volume 338.
\newblock Springer.

\end{thebibliography}

\ifbool{finalsubmission}{}{
\appendix
\section{Proofs}
\subsection{Proof of Proposition~\ref{prop:projection}}
\begin{pf}
We split the proof in three parts.\\
\emph{Existence:} Since the Wasserstein is lower semi-continuous (w.r.t. narrow convergence), and has compact (w.r.t. narrow convergence) level sets, it suffices to prove that the set $\{\nu\in\Pp{}{\reals}:\support{\nu}\subset\Theta\}\subset\Pp{}{\reals^d}$ is closed (w.r.t. narrow convergence). Let $(\nu_n)_{n\in\naturals}\subset\Pp{}{\reals^d}$ so that $\support{\nu_n}\subset\Theta$ for all $n\in\naturals$, and assume that $(\nu_n)_{n\in\naturals}$ converges narrowly to $\nu\in\Pp{}{\reals^d}$. We seek to prove that $\support{\nu}\subset\Theta$.
For $m\in\naturals$, let $f_m(x)$ be continuous and bounded so that (i) $f_m(x)=1$ if $x\in\Theta$ and (ii) converges pointwise to $1_\Theta(x)$ as $m\to\infty$. As $\Theta$ is closed, such $f_m$ can always be constructed. 
Since $\support{\nu_n}\subset\Theta$ for every $m\in\naturals$ and every $n\in\naturals$
\begin{equation*}
    \int_{\reals^d}f_m(x)\d\nu_n(x)=\int_{\Theta}f_m(x)\d\nu_n(x)=\nu_n(\Theta)=1. 
\end{equation*}
Moreover, as $f_m$ is continuous and bounded, for any fixed $m\in\naturals$ the definition of narrow convergence gives
\begin{equation*}
    1=\lim_{n\to\infty}\int_{\reals^d}f_m(x)\d\nu_n(x)=\int_{\reals^d}f_m(x)\d\nu(x).
\end{equation*}
We can now deploy dominated convergence ($|f_m|$ is uniformly dominated by an $\nu$-integrable function) to conclude
\begin{equation*}
\begin{aligned}
    1
    =
    \lim_{m\to\infty}\int_{\reals^d}f_m(x)\d\nu(x)
    &=
    \int_{\reals^d}\lim_{m\to\infty}f_m(x)\d\nu(x)
    \\
    &=
    \int_{\reals^d}1_\mathrm{\Theta}(x)\d\nu(x) =\nu(\Theta).
\end{aligned}
\end{equation*}
Thus, $\support{\nu}\subset\Theta$, and the set $\{\nu\in\Pp{}{\reals^d}:\support{\nu}\subset\Theta\}$ is closed (w.r.t. narrow convergence), as desired. \\
\emph{Uniqueness:} Assume two non-equal projections $\bar\mu_1,\bar\mu_2\in\Pp{2}{\reals^d}$ exist. Let $\gamma_1\in\setOptimalPlans{\bar\mu}{\bar\mu_1}$ and $\gamma_2\in\setOptimalPlans{\bar\mu}{\bar\mu_2}$, and let $\mu_{1/2}$ be the generalized geodesics with $\mu_{1/2}=\pushforward{\left(\frac{1}{2}\proj_2+\frac{1}{2}\proj_3\right)}\gamma$,
where $\gamma\in\setPlans{\bar\mu,\bar\mu_1}{\bar\mu_2}\subset\Pp{}{\reals^d\times\reals^d\times\reals^\d}$ results from gluing $\gamma_1$ and $\gamma_2$ (via the Gluing Lemma, e.g., Lemma 5.3.2 in~\citet{ambrosio2005gradient}; more generally, see Chapter 9.2 in~\citet{ambrosio2005gradient} for an introduction to generalized geodesics. 
Since $\gamma_1$ and $\gamma_2$ are both concentrated on $\Theta\times\Theta$, $\gamma$ must also have support in $\Theta\times\Theta\times\Theta$. Thus,  
\begin{equation*}
\begin{aligned}
\support{\mu_{1/2}}
&\subset\left(\frac{1}{2}\proj_2+\frac{1}{2}\proj_3\right)(\Theta\times\Theta\times\Theta)
\\
&=
\frac{1}{2}\Theta\oplus\frac{1}{2}\Theta = \Theta,
\end{aligned}
\end{equation*}
where the last equality follows from the convexity of $\Theta$ and the definition of Minkovsky sum of sets. This shows that $\mu_{1/2}$ is feasible. 
The squared Wasserstein distance from $\mu$ is known to be 2-convex along this geodesic. Thus,
\begin{equation*}
\begin{aligned}
    \wassersteinDistance{2}{\mu_{1/2}}{\bar\mu}^2
    &\leq 
    \frac{1}{2}\wassersteinDistance{2}{\bar\mu_{0}}{\bar\mu}^2\!+\!\frac{1}{2}\wassersteinDistance{2}{\mu_{1}}{\bar\mu}\!-\!\frac{1}{4}\wassersteinDistance{2}{\bar\mu_0}{\bar\mu_1}^2
    \\
    &=\wassersteinDistance{2}{\bar\mu_{0}}{\bar\mu}^2-\frac{1}{4}\wassersteinDistance{2}{\bar\mu_0}{\bar\mu_1}^2.
\end{aligned}
\end{equation*}
Since $\bar\mu_0\neq\bar\mu_1$, $\wassersteinDistance{2}{\bar\mu_0}{\bar\mu_1}>0$. 
However, this implies $\wassersteinDistance{2}{\mu_{1/2}}{\bar\mu}<\wassersteinDistance{2}{\bar\mu_{0}}{\bar\mu}$, which contradicts optimality of $\bar\mu_{0}$ and $\bar\mu_1$.\\
\emph{Equation \eqref{eq:projection particles}:} We will use Kantorovich duality. Let $\nu\in\Pp{2}{\reals^d}$ with $\support{\nu}\subset\Theta$ and let $1_\Theta(x)$ be 0 in $\Omega$ and $+\infty$ outside. Clearly, $1_\Theta$ is zero $\nu$-a.e., and so $\int_{\reals^d}1_{\Theta}(x)\d\nu(x)=0$ for any $\nu\in\mathcal{A}$. Thus, Kantorovich duality (e.g., Chapter 5 in~\cite{villani2009optimal}), with $f^{c}(y)=\sup_{x\in\reals^d}f(x)-\norm{x-y}^2$, gives 
\begin{equation}\label{eq:proof projection lb}
\begin{aligned}
    \wassersteinDistance{2}{\nu}{\mu}^2
    &\geq 
    \int_{\reals^d}1_{\Theta}(x)\d\nu(x)-\int_{\reals^d}(1_\Theta)^{c}(y)\d\mu(y)
    \\
    &=
    -\int_{\reals^d}\sup_{x\in\reals^d}1_\Theta(x)-\norm{x-y}^2\d\mu(x)
    \\
    &=
    \int_{\reals^d}\inf_{x\in\Theta}\norm{x-y}^2\d\mu(x).
\end{aligned}
\end{equation}
Moreover, $\proj_\Theta[\cdot]$ is trivially a (possibly suboptimal) transport map from $\mu$ to $\pushforward{(\proj_\Theta[\cdot])}\mu$. Thus, 
\begin{equation}\label{eq:proof projection ub}
\begin{aligned}
    \wassersteinDistance{2}{\pushforward{(\proj_\Theta[\cdot])}\mu}{\mu}^2
    &\leq 
    \int_{\reals^d}\norm{x-\proj_{\Theta}[x]}^2\d\mu(x)
    \\
    &=
    \int_{\reals^d}\inf_{x\in\Theta}\norm{x-y}^2\d\mu(x),
\end{aligned}
\end{equation}
where the last equality follows from the definition of projection. We can now combine~\eqref{eq:proof projection lb} and~\eqref{eq:proof projection ub} to conclude that for all $\nu\in\Pp{2}{\reals^d}$ with $\support{\nu}\subset\Theta$
\begin{equation*}
    \wassersteinDistance{2}{\nu}{\mu}
    \geq 
    \wassersteinDistance{2}{\pushforward{(\proj_\Theta[\cdot])}\mu}{\mu}.
\end{equation*}
Since $\support{{\pushforward{(\proj_\Theta[\cdot])}\mu}}\subset\Theta$ and projections are unique, we establish~\eqref{eq:projection particles}. This concludes the proof. 
\end{pf}

\subsection{Proof of Theorem~\ref{thm:convergence}}
\begin{pf}
We start with the proof of~\eqref{eq:theorem convergence finite}. The other statements then follow. 
Let $\gamma\in\setOptimalPlans{\mu_k}{\mu^\ast}$, where $\mu^\ast$ is well-defined; indeed, if $J$ is $\alpha$-geodesically convex with $\alpha>0$ and lower semi-continuous w.r.t. the convergence induced by the Wasserstein distance, a unique minimizer exists (e.g., see Section 11.2 in~\cite{ambrosio2005gradient}). 
Then, 
\begin{align*}
    &\expectedValue{\xi_k}{\wassersteinDistance{2}{\mu_{k+1}}{\mu^\ast}^2}
    \\
    \overset{\clubsuit}&{=}
    \expectedValue{\xi_k}{\wassersteinDistance{2}{\pushforward{(\proj_{\Theta})\circ(\Id-\tau\xi_k)}\mu_k}{\mu^\ast}^2}
    \\
    \overset{\heartsuit}&{\leq} 
    \begin{aligned}[t]
    \mathbb{E}^{\xi_k}&\left[\int_{\reals^d}\norm{\theta_k-\theta^\ast}^2\right. \\
    &\qquad\left.\d\left(\pushforward{\left(\proj_\Theta\circ(\Id-\tau\xi_k))\times\Id\right)}\gamma\right)(\theta_k,\theta^\ast)\right]
    \end{aligned}
    \\
    \overset{\square}&{=}
    \expectedValue{\xi_k}{\int_{\reals^d}\norm{\proj_{\Theta}[\theta_k-\tau\xi_k(\theta_k)]-\theta^\ast}^2\d\gamma(\theta_k,\theta^\ast)}
    \\
    \overset{\triangle}&{\leq}
    \expectedValue{\xi_k}{\int_{\reals^d}\norm{\theta_k-\tau\xi_k(\theta_k)-\theta^\ast}^2\d\gamma(\theta_k,\theta^\ast)}
    \\
    &=
    \begin{aligned}[t]
    &\expectedValue{\xi_k}{\int_{\reals^d}\norm{\theta_k-\theta^\ast}^2\d\gamma(\theta_k,\theta^\ast)}\\
    &+\tau^2\expectedValue{\xi_k}{\int_{\reals^d}\norm{\xi_k(\theta_k)}^2\d\gamma(\theta_k,\theta^\ast)} \\
    &-2\tau\expectedValue{\xi_k}{\int_{\reals^d}\innerProduct{\theta_k-\theta^\ast}{\xi_k(\theta_k)}\d\gamma(\theta_k,\theta^\ast)}
    \end{aligned}
    \\
    \overset{\spadesuit}&{\leq}
    \begin{aligned}[t]
    &\wassersteinDistance{2}{\mu_k}{\mu^\ast}^2
    +\tau^2\expectedValue{\xi_k}{\int_{\reals^d}\norm{\xi_k(\theta_k)}^2\d\mu_k(\theta_k)} \\
    &-2\tau\expectedValue{\xi_k}{\int_{\reals^d}\innerProduct{\theta_k-\theta^\ast}{\xi_k(\theta_k)}\d\gamma(\theta_k,\theta^\ast)}
    \end{aligned}
    \\
    &\leq
    \begin{aligned}[t]
    &\wassersteinDistance{2}{\mu_k}{\mu^\ast}^2
    +\tau^2\expectedValue{\xi_k}{\int_{\reals^d}\norm{\xi_k(\theta_k)}^2\d\mu_k(\theta_k)} \\
    &-2\tau\int_{\reals^d}\innerProduct{\theta_k-\theta^\ast}{\expectedValue{\xi_k}{\xi_k(\theta_k)}}\d\gamma(\theta_k,\theta^\ast)
    \end{aligned}
    \\
    &\leq
    \begin{aligned}[t]
    &\wassersteinDistance{2}{\mu_k}{\mu^\ast}^2
    +\tau^2\expectedValue{\xi_k}{\int_{\reals^d}\norm{\xi_k(\theta_k)}^2\d\mu_k(\theta_k)} \\
    &-2\tau\int_{\reals^d}\innerProduct{\theta_k-\theta^\ast}{\gradient{\mu} J(\mu)(\theta_k)}\d\gamma(\theta_k,\theta^\ast)
    \end{aligned}
    \\
    \overset{\diamondsuit}&{\leq}
    \begin{aligned}[t]
    &\wassersteinDistance{2}{\mu_k}{\mu^\ast}^2+\tau^2(\sigma^2+C(J(\mu_k)-J(\mu^\ast))) \\&+2\tau\left(J(\mu^\ast)-J(\mu_k)-\frac{\alpha}{2}\wassersteinDistance{2}{\mu_k}{\mu^\ast}^2\right)
    \end{aligned}
    \\
    &\leq 
    \begin{aligned}[t]
    &(1-\alpha\tau)\wassersteinDistance{2}{\mu_k}{\mu^\ast}^2 + \tau^2\sigma^2 \\ &+(J(\mu_k)-J(\mu^\ast))(C\tau^2-2\tau),
    \end{aligned}
\end{align*}
where in $\clubsuit$ we used the definition of $\mu_{k+1}$; in $\heartsuit$ we used that $\pushforward{(\proj_\Theta\circ(\Id-\tau\xi_k)\times\Id)}\gamma$ is a (possibly sub-optimal) transport plan between $\pushforward{(\proj_{\Theta}\circ(\Id-\tau\xi_k))}\mu_k$ and $\mu^\ast$~(by Lemma 3.3~\citet{aolaritei2022uncertainty}), i.e., 
\begin{equation*}
    \pushforward{(\proj_\Theta\circ(\Id\!-\tau\xi_k)\times\Id)}\gamma
    \in
    \setPlans{\pushforward{(\proj_{\Theta}\!\circ(\Id\!-\tau\xi_k))}\mu_k}{\mu^\ast}
\end{equation*}
is candidate (but generally suboptimal) plan for the Wasserstein distance between $\pushforward{(\proj_{\Theta}\circ(\Id-\tau\xi_k))}\mu_k$ and $\mu^\ast$; in $\square$ we used $\int g\d\pushforward{f}\mu=\int g\circ f\d\mu$; in $\triangle$ we used non-expansivness of the projection operator (together with $\proj_{\Theta}[\theta^\ast]=\theta^\ast$ for all $\theta^\ast\in\support{\mu^\ast}\subset\Theta$); $\spadesuit$ follows from the definition of $\gamma$; and in $\diamondsuit$ we used properties of Wasserstein gradients of $\alpha$-convex functionals (e.g., see Proposition 2.8 in~\cite{lanzetti2022first}), together with Assumption~\ref{ass:finite variance}.
Since by assumption $\tau\leq 2/C$, $C\tau^2-2\tau\leq 0$, and 
\begin{equation*}\label{eq:proof onestep}
    \expectedValue{\xi_k}{\wassersteinDistance{2}{\mu_{k+1}}{\mu^\ast}^2}
    \leq 
    (1-\alpha\tau)\wassersteinDistance{2}{\mu_k}{\mu^\ast}^2 + \tau^2\sigma^2.
\end{equation*}
We can now proceed iteratively to obtain
\begin{equation*}
\begin{aligned}
    &\expectedValue{\{\xi_j\}_{j=0}^k}{\wassersteinDistance{2}{\mu_{k+1}}{\mu^\ast}^2}
    \\
    &\leq
    (1-\tau\alpha)^k\left(\wassersteinDistance{2}{\mu_0}{\mu^\ast}^2-\frac{\sigma^2\tau}{\alpha}\right)+\frac{\sigma^2\tau}{\alpha},
\end{aligned}
\end{equation*}
This establishes~\eqref{eq:theorem convergence finite}.\\
We now prove~\eqref{eq:theorem convergence asymptotic}. By assumption $0<\tau<1/\alpha$, $1-\alpha\tau\in(0,1)$, and so the limit $k\to\infty$,~\eqref{eq:theorem convergence finite} gives
\begin{equation*}
    \limsup_{k\to\infty}\expectedValue{\{\xi_j\}_{j=0}^k}{\wassersteinDistance{2}{\mu_{k+1}}{\mu^\ast}^2}
    \leq 
    \frac{\tau\sigma^2}{\alpha}.
\end{equation*}
By Jensen inequality, together with continuity and monotonicity of $x\mapsto x^2$, we have
\begin{equation*}
\begin{aligned}
    &\hspace{-1cm}\left(\limsup_{k\to\infty}\expectedValue{\{\xi_j\}_{j=0}^k}{\wassersteinDistance{2}{\mu_{k+1}}{\mu^\ast}}\right)^2
    \\
    &=
    \limsup_{k\to\infty}\left(\expectedValue{\{\xi_j\}_{j=0}^k}{\wassersteinDistance{2}{\mu_{k+1}}{\mu^\ast}}\right)^2
    \\
    &\leq 
    \limsup_{k\to\infty}\expectedValue{\{\xi_j\}_{j=0}^k}{\wassersteinDistance{2}{\mu_{k+1}}{\mu^\ast}^2} \leq
    \frac{\tau\sigma^2}{\alpha}.
\end{aligned}
\end{equation*}
Monotonicity $x\mapsto\sqrt{x}$ establishes~\eqref{eq:theorem convergence asymptotic}. 
\\
We now prove~\eqref{eq:theorem convergence mean}, it suffices to observe that, in virtue of Gelbrich's bound (\cite{gelbrich1990formula}), we have
\begin{equation*}
    \wassersteinDistance{2}{\mu}{\nu}
    \geq 
    \sqrt{\norm{m_{\mu}-m_{\nu}}^2+d(S_{\mu},S_{\nu})^2}
\end{equation*}
with $m_\mu$ and and $S_{\mu}$ ($m_\nu$ and $S_{\nu}$) being the mean and covariance matrices of $\mu$ ($\nu$), and $d$ being the Bures distance on symmetric positive semidefinite matrices,  defined in~\eqref{eq:bures metric}.
Thus,~\eqref{eq:theorem convergence mean} follows from~\eqref{eq:theorem convergence asymptotic}.
\\
Finally,~\eqref{eq:theorem convergence lipschitz} follows from \citet[Proposition 7.29]{villani2009optimal}, observing that $\reals^d$ is locally compact and replacing $\varphi$ via $|\varphi|$.
This concludes the proof.
\end{pf}

\subsection{Proof of Corollary~\ref{cor:noise free}}
\begin{pf}
The proof follows directly from Theorem~\ref{thm:convergence}, together with the well-known fact convergence in the Wasserstein distance is equivalent to weak convergence in $\Pp{2}{\reals^d}$; see Chapter 6 in~\cite{villani2009optimal}.
\end{pf}

\subsection{Proof of Lemma~\ref{lemma:gradient}}
\begin{pf}
We prove the statements separately.
\\ 
The proof of~\eqref{eq:wasserstein gradient example} follows from Section 2 in~\cite{lanzetti2022first}.
To prove~\eqref{eq:wasserstein gradient unbiased} observe
\begin{equation*}
\begin{aligned}
    \expectedValue{}{\xi_k}
    &=
    \expectedValue{}{W^\top \left(W\theta -\hat y_i\right)+\rho\left(\theta-\expectedValue{\mu}{\theta}\right)}
    \\
    &=
    W^\top \left(W\theta-\expectedValue{}{\hat y_i}\right)
    +
    \rho\left(\theta-\expectedValue{\mu}{\theta}\right)
    \\
    &=
    W^\top
    \left(W\theta-W\theta^\ast\right)
    +
    \rho\left(\theta-\expectedValue{\mu}{\theta}\right)
    =
    \gradient{\mu}J(\mu).
\end{aligned}
\end{equation*}
\\ 
For the proof of~\eqref{eq:wasserstein gradient finite variance}, observe that 
\begin{equation*}
\begin{aligned}
    &\expectedValue{}{\LpNorm{\xi}{2}{\reals^d,\reals^d;\mu}^2}
    \\
    &=
    \expectedValue{}{\LpNorm{W^\top(W\theta-W\theta^\ast-w) + \rho\left(\theta-\expectedValue{\mu}{\theta}\right)}{2}{\reals^d,\reals^d;\mu}^2}
    \\
    &\leq
    \begin{aligned}[t]
    &2\expectedValue{}{\int_{\Theta}\norm{W^\top \left(W\theta-W\theta^\ast-w\right)}^2\d\mu(\theta)}
    \\
    &+2\rho^2\int_{\Theta}\norm{\theta-\expectedValue{\mu}{\theta}}^2\d\mu(\theta)
    \end{aligned}
    \\
    &\leq
    \begin{aligned}[t]
    &4\sigma_\mathrm{max}(W)^2\frac{1}{2}\expectedValue{}{\int_{\Theta}\norm{ W\theta-W\theta^\ast-w}^2\d\mu(\theta)}
    \\
    &+4\rho\frac{\rho}{2}\int_{\Theta}\norm{\theta-\expectedValue{\mu}{\theta}}^2\d\mu(\theta)
    \end{aligned}
    \\
    &\leq 
    4\max\{\sigma_\mathrm{max}(W)^2,\rho\} J(\mu),
\end{aligned}
\end{equation*}
where we used the definition of $J$ for the last inequality.
Thus,
\begin{equation*}
\begin{aligned}
    &\expectedValue{}{\LpNorm{\xi}{2}{\reals^d,\reals^d;\mu}^2}
    \\
    &\leq 
    4\max\{\sigma_\mathrm{max}(W)^2,\rho\} \left(J(\mu)-J(\mu^\ast) + J(\mu^\ast)\right).
    \\
    &\leq 
    4\max\{\sigma_\mathrm{max}(W)^2,\rho\} \left(J(\mu)-J(\mu^\ast) + \int_{\mathcal W}\norm{w}^2\d\nu(w)\right)
    \\
    &=
    \begin{aligned}[t]
    &4\max\{\sigma_\mathrm{max}(W)^2,\rho\} \left(J(\mu)-J(\mu^\ast)\right) \\
    &+ 4\max\{\sigma_\mathrm{max}(W)^2,\rho\}\sigma^2.
    \end{aligned}
\end{aligned}
\end{equation*}
This concludes the proof.
\end{pf}

\subsection{Proof of Corollary~\ref{cor:convergence example}}
\begin{pf}
We just need to evaluate the convexity parameters of $J$ and $\sigma^2, C$, as defined in Assumption~\ref{ass:finite variance}. Since $\mu\mapsto\expectedValue{\mu}{f}$ is $\alpha$-geodesically convex if and only if $f$ is $\alpha$-convex, we have that $J$ is geodesically convex with $\alpha=\lambda_\mathrm{min}(\transpose{W}W)=\sigma_\mathrm{min}(W)^2$. Moreover, by Lemma~\ref{lemma:gradient}, we have 
\begin{equation*}
\begin{aligned}
    \sigma^2 &=4\max\{\sigma_\mathrm{max}(W)^2,\rho\}\sigma_w^2
    \\
    C &=4\max\{\sigma_\mathrm{max}(W)^2,\rho\}.
\end{aligned}
\end{equation*}
Then, the result follows from Theorem~\ref{thm:convergence}, as
\begin{equation*}
    \frac{\sigma^2\tau}{\alpha}=\frac{4\max\{\sigma_\mathrm{max}(W)^2,\rho\}\sigma_w^2\tau}{\sigma_\mathrm{min}(W)^2}=\eta\sigma_w^2\tau.
\end{equation*}
Also, $\alpha\leq 2C$. Thus, the condition on $\tau$ simplifies to $\tau\in (0,1/(2\max\{\sigma_\mathrm{max}(W)^2,\rho\}))$. This concludes the proof. 
\end{pf}}

\end{document}